\documentstyle[11pt]{article}
\input{epsf}
\begin{document}

\title{ Inner edge of neutron-star crust with SLy\\
        effective  nucleon-nucleon interactions}

\author{{\bf F.\ Douchin}$^{1}$, {\bf P.\ Haensel}$^{1,2,3}$\\
        $^1${\it Centre de Recherche Astronomique de Lyon, } \\
        {\it Ecole Normale Sup{\'e}rieure de Lyon,
          46, all{\'e}e d'Italie,}\\
        {\it 69364 Lyon,  France}\\
        $^2${\it N.\ Copernicus Astronomical Center,} \\
        {\it Polish Academy of Sciences, Bartycka 18,}\\
        {\it 00-716 Warszawa, Poland}\\
        $^3${\it D{\'e}partement d'Astrophysique Relativiste et de 
           Cosmologie, }\\{\it  UMR 8629 du CNRS,
         Observatoire de Paris,}\\
          {\it  F-92195 Meudon Cedex, France}\\
        }

\maketitle
\vskip 3mm
\centerline{\sl  to be published in Physics Letters B}
\vskip 3mm


\def\la{\;\raise0.3ex\hbox{$<$\kern-0.75em\raise-1.1ex\hbox{$\sim$}}\;}
\def\ga{\;\raise0.3ex\hbox{$>$\kern-0.75em\raise-1.1ex\hbox{$\sim$}}\;}
\newcommand{\dd}{\mbox{d}}                     
\begin{abstract}
The boundary between the solid crust, assumed to be in complete
thermodynamic equilibrium (cold catalyzed matter),
 and the liquid core of a neutron star
 is studied using   Skyrme
SLy effective N-N interactions.
An approximate value of the density at the inner edge of the
 crust is obtained from the threshold
for the instability of  homogeneous $npe$ matter with respect to
small periodic density perturbations. Calculations of the bottom
layer of the crust are  performed within the Compressible Liquid
Drop Model.  Spherical nuclei are energetically preferred over
exotic ones (cylinders, plates, tubes, bubbles), down to the inner
edge of the crust, found at $\rho_{\rm edge}= 0.08~{\rm fm^{-3}}$.
\end{abstract}
\vskip 3mm

 {\it Key words}: Dense matter. Neutron stars.

 PACS numbers: 97.60.Jd, 21.65.+f, 95.30.Cq
\par
\vskip 1cm
The solid crust of a neutron star plays an important role in neutron star
evolution and dynamics. It insulates thermally neutron star surface from
its hot liquid interior, and therefore plays an essential role in neutron
star cooling. It can build-up stresses, leading to instabilities responsible
for the glitches in pulsar timing. Moreover, it  can support non-axial
deformations
which, combined with rapid rotation, could make  neutron star  a source
of continuous gravitational radiation.

The mass of the neutron-star crust depends sensitively on the
density of its inner edge and on the neutron-star matter equation
of state. Recent calculations of this quantity
 \cite{LRP93,Oyam93,PethRL95,Cheng97}
give $\rho_{\rm edge}\simeq 0.08-0.10~{\rm fm^{-3}}$,
substantially lower than the estimate close to the normal nuclear
density $\rho_0=0.16~{\rm fm^{-3}}$, derived in the classical
paper of Baym et al. \cite{BBP}.
 Some  theoretical calculations predict exotic distributions of nuclear matter
 (rods, plates, etc.) within the bottom layers  of the
crust \cite{LRP93,Oyam93,Sumiy95}. However, the presence of exotic
nuclei depends on the effective nucleon-nucleon (N-N) force used;
for some forces, spherical nuclei are present down
 to the inner edge of the crust \cite{LRP93,Cheng97}.
Presence of exotic nuclei would lead to different
 transport and  elastic properties of the crust, compared to the
standard case of a Coulomb crystal formed by spherical nuclei
\cite{LRP93, PethPot98}. This, in turn, would have important
consequences for  dynamics and evolution of neutron star.

In order to make  calculations of the structure of the bottom
layers of the neutron-star crust feasible, one has to use an
effective N-N interaction. Frequently used in astrophysical
applications SkM$^*$ force was constructed to provide a consistent
description of isovector properties of nuclei (giant dipole
resonances) \cite{Bartel.etal82}, and seems thus to be
particularly suitable for  the description of neutron rich nuclei.
However,  this as well as many other existing effective N-N
interactions were  fitted to the properties of laboratory atomic
nuclei, with $(N-Z)/A < 0.3$,
 while in the bottom layers of neutron-star crust one expects
$(\rho_n-\rho_p)/\rho\ga 0.8$. In view of this, application of
these effective N-N interactions to the bottom layers of a
neutron-star crust involves a rather risky extrapolation to
strongly asymmetric nucleon matter. In order to remove a part of
this uncertainty, modifications of  effective N-N  forces, to make
them consistent with microscopic calculations of neutron matter,
have been applied. Such a procedure was used in  seventies to
obtain the Sk$1^\prime$ force \cite{LattRav78}, via a rather {\it
ad hoc} modification of the Sk1 force constructed originally by
Vautherin and Brink \cite{VauthB70} to describe terrestrial
nuclei. In this way, it became consistent with energy per nucleon
of neutron matter calculated by Siemens and Pandharipande
\cite{SiemPand71}. Later, generalized types of the Skyrme
interaction, FPS \cite{PandRavFPS89} and FPS21 \cite{PethRL95},
with larger number of fitted parameters and more general density
dependence, were derived by fitting  the temperature and density
dependent energies per baryon of nuclear and neutron matter
obtained in microscopic calculations of Friedman and Pandharipande
\cite{FP81}.

Recently, a new  set of the Skyrme-type effective N-N interaction has
been derived, based on an approach which may be more appropriate, as
far as the applications to a very neutron rich matter are concerned
\cite{Chabanat1,Chabanat2}. Relevant additional experimental items
concerning neutron rich nuclei (isovector effective
masses), constraints of spin stability, and requirement of consistency
with the UV14+VIII equation of state (EOS)
  of dense neutron matter of Wiringa et al. \cite{WFF88}
  for $\rho_0\le \rho\le 1.5~{\rm fm^{-3}}$, were combined with
general procedure of fitting the properties of doubly
 magic nuclei. This procedure led to a set of the SLy (Skyrme Lyon) models,
 which - due to the emphasis
 put on their neutron-excess dependence - seem to be particularly
suitable for the calculations of the properties of neutron-star
crust.

In the present Letter,  we calculate  the structure of the bottom
layer of the neutron-star crust, and the density at its inner
edge, $\rho_{\rm edge}$,
 using  the SLy models of effective
N-N interaction, and compare our results with those obtained using
older Skyrme-type
forces,  SkM$^*$ and Sk1$^\prime$.
  The parameters of the SLy forces
used in the present calculations, together with those of the
SkM$^*$ and Sk1$^\prime$ models, are given in Table 1. The SLy4
interaction is a basic SLy force. The   SLy7 model has been
obtained
 following the most ambitious fitting procedure, in which
spin-gradient
terms and center of mass correction term  were both included in the Skyrme
energy functional \cite{Chabanat2}.

Recent calculations show that neutron-star matter  (both liquid
and solid) close to the crust-liquid core interface contains only
a few percent of protons. Therefore, an  effective N-N interaction
used for the calculation of the crust-liquid transition should, in
the first place, yield a realistic description of pure neutron
matter at subnuclear density.  As  we do not have a direct access
to the EOS of pure neutron matter at subnuclear densities, results
of
 precise numerical calculations
of the ground state at  $\rho<\rho_0$,  carried out using modern
many-body theories  and the best {\it bare} N-N hamiltonians,  are
to be used as an ersatz  of such  experimental data
\cite{PethRL95}.

The SLy forces have been constructed as to be consistent with the
UV14+UVII  model of Wiringa et al. \cite{WFF88} of neutron matter
above $\rho_0$
 \cite{Chabanat1,Chabanat2}.
It is therefore of interest to see how well these effective N-N
forces reproduce the  UV14+UVII model of neutron matter at
subnuclear densities.

In Fig. 1 we have plotted the energy per neutron, $e$, versus
neutron density, $\rho<\rho_0$,  for the SLy4 and SLy7 effective
N-N interactions. The filled squares correspond to realistic
UV14+UVII model of neutron matter, and were taken from Table III
of Wiringa et al. \cite{WFF88}.
 The agreement of the SLy4, SLy7  curves with the neutron matter EOS of  the
UV14+VII model of Wiringa et al. \cite{WFF88} is very good. This is
not the case for older Skyrme forces, Sk$1^\prime$ and SkM$^*$.

In what follows, a mixture of neutrons, protons, and electrons
will be referred to as $npe$ matter. An electrically neutral $npe$
matter in beta equilibrium corresponds to neutron-star matter at
$\rho\sim \rho_0$.  At a given $\rho$, the ground state of a
homogeneous $npe$ matter
 corresponds to the minimum
of the energy density $E(\rho_n,\rho_p,\rho_e)=E_0$,
 under the constraints of  fixed baryon density  and electric charge
neutrality,
$\rho_p+\rho_n=\rho$
and $\rho_e=\rho_p$, respectively.
 This  implies beta equilibrium between the matter constituents and
ensures  vanishing of the first  variation of  $E$ due to
 small perturbations $\delta\rho_{j}({\bf r})$
(where $j=n,p,e$) of the equilibrium
 solution (under the constraints of constant total nucleon number,
$A=\rho V$,  and
global charge neutrality within the volume $V$ of the system). However, this
 does not guarantee the stability of the spatially
homogeneous state of the $npe$ matter, which requires that the
second variation  of $E$ (quadratic in $\delta\rho_j$)
 be positive.

Our expression for the energy functional of slightly inhomogeneous
neutron-star matter  has been calculated using the semi-classical
Extended Thomas-Fermi (ETF) treatment  of the kinetic  and
 the spin-gradient terms in nucleon contribution to $E$
  \cite{BrackJC76}.
Assuming that the spatial gradients are small, we  keep only the
quadratic gradient terms in the ETF expressions. Our approximation
is  justified by the fact that characteristic wavelengths of
periodic perturbations turn out to
 be much larger than the
internucleon distance.
 With these approximations,
 the change of the energy (per unit volume)
 implied
by the  density perturbations can be expressed, keeping only second order
terms \cite{BBP,PethRL95},
\begin{equation}
E-E_0= {1\over 2}\int{{\rm d}{\bf q}\over (2\pi)^3}
\sum_{i,j} F_{ij}({\bf q})\delta\rho_i({\bf q})\delta\rho_j({\bf q})^*~,
\label{E-E_0}
\end{equation}
where we used the Fourier representation
\begin{equation}
\delta\rho_j({\bf r})=\int{{\rm d}{\bf q}\over (2\pi)^3}
\delta\rho_j({\bf q}){\rm e}^{{\rm i}{\bf q}{\bf r}}~.
\label{rho.Fourier}
\end{equation}
The hermitian
$F_{ij}({\bf q})$
matrix determines the
stability of the uniform state
of equilibrium of the $npe$ matter with respect to the spatially periodic
perturbations  of wavevector ${\bf q}$.
Due to the isotropy of the homogeneous equilibrium state of
the $npe$ matter, $F_{ij}$ depends only on $\vert {\bf q}\vert=q$.
In the case of Skyrme-type forces,
 the matrix elements
$F_{ij}$ can be calculated analytically, as explicit functions
of the equilibrium densities and $q$, and are composed of compression
(local), curvature (gradient), and Coulomb components \cite{BBP,PethRL95}.

The condition for the $F_{ij}$ matrix to be positive definite is
 equivalent to the requirement that the determinant of the $F_{ij}$
matrix be positive \cite{PethRL95}. At each density $\rho$,
one has thus to check the
condition
${\rm det}[F_{ij}(q)]>0$.
 Let us start with some $\rho$, at which
${\rm det}[F_{ij}(q)]>0$ for any $q$. By decreasing $\rho$, we
find eventually a wavenumber $Q$ at which stability condition is
violated for the first time; this happens at some density
$\rho_Q$. For $\rho<\rho_Q$ the homogeneous state is no longer the
ground state of the $npe$ matter since it is  unstable with
respect to small periodic density modulations.

We get $\rho_Q=0.079~{\rm fm^{-3}}$ for both SLy4 and SLy7 forces,
to be compared with $0.075~{\rm fm^{-3}}$ and $0.101~{\rm
fm^{-3}}$ for the SkM$^*$ and Sk$1^\prime$ forces, respectively.
 The instability at $\rho_Q$
signals  a  phase  transition with a loss of translational
symmetry of the $npe$ matter, and appearance of nuclear
structures. Calculations performed previously with  Skyrme forces
indicate that $\rho_Q\simeq \rho_{\rm edge}$ \cite{PethRL95}.
The differences between the values of $\rho_Q$ between four 
forces are correlated with differences in the density behavior of 
nucleon chemical potentials, whose density dervatives enter the 
stability matrix $F_{ij}$ \cite{PethRL95}. In Fig.2 we plot the 
values of $\mu_n$ and $\mu_p$ for all four Skyrme forces considered, 
versus nucleon density $\rho$, for the $npe$ matter in beta 
equilibrium. In the relevant density region, $\rho\sim {1\over 2} 
\rho_0$, one notices very good agreement between the values of 
$\mu_p$. The relative differences in the values of $\mu_n$, 
and in particular in the slopes of the $\mu_n(\rho)$ curves are 
much larger. The slope of $\mu_n(\rho)$ for the SkM$^*$ force is the 
highest, and that for the Sk1$^\prime$ force the lowest, at $\rho\simeq 
{1\over 2}\rho_0$. Notice that the slope of $\mu_n$ for the Sk1$^\prime$ 
is quite similar to those obtained  in \cite{PethRL95} for the FPS and 
FPS21 forces (see Fig.5 of \cite{PethRL95}). Qualitatively, the higher the 
slope of $\mu_n$, the lower the value of $\rho_Q$ (c.f., \cite{PethRL95}).

Nuclear structures in bottom layers of neutron-star crust were
described using the Compressible Liquid Drop Model (CLDM) of
nuclei \cite{BBP,LPRL85,LorenzThesis,DouchinThesis}. Within the
CLDM, one is able to separate bulk, surface, and Coulomb
contributions to total energy density, $E$. Electrons are assumed
to form an uniform Fermi gas, and yield the rest and kinetic
energy contribution, denoted by $E_e$. Within CLDM, total energy
density of the inner neutron-star crust is given by
\begin{equation}
E=E_{{\rm N,bulk}} + E_{{\rm N,surf}} + E_{\rm Coul} + E_{e}~.
\label{E.CLDM}
\end{equation}
Here, $E_{{\rm N,bulk}}$ is the bulk contribution of nucleons,
which does not depend on the shape of nuclear structures. However,
both $E_{{\rm N,surf}}$ and  $E_{\rm Coul}$ do depend on the shape
of nuclear structures, formed by denser nuclear matter and  the
less dense neutron gas (detailed description of the calculation of
$E_{\rm N,surf}$ and $E_{\rm Coul}$ for the SLy forces was
presented
 in \cite{DouchinThesis}; our results for the nuclear surface
and curvature properties for  SLy forces will be described  in
detail elsewhere \cite{Esurf}). We restricted ourselves to three
shapes of the nuclear matter - neutron gas interface: spherical,
cylindrical, and plane. Consequently, we considered five types of
nuclear structures: spheres of nuclear matter in neutron gas,
cylinders of nuclear matter in neutron gas (rods), plane slabs of
nuclear matter in neutron gas, cylindrical holes in nuclear matter
filled by neutron gas (tubes), and spherical holes in nuclear
matter filled by neutron gas (bubbles). In view of a significant
neutron excess, the interface includes neutron skin, formed by
neutrons adsorbed onto the nuclear matter surface. In view of a
finite thickness of nuclear surface, the definition of its spatial
location is a matter of convention. Here, we defined it by the
radius of the equivalent constant density  proton 
distribution, $R_p$, which
determines thus the radius of spheres, bubbles, cylinders and
tubes, and the half-thickness of plane nuclear matter slabs. 
The neutron radius, $R_n$, 
was defined by the condition that it yields a squared-off neutron 
density ditribution with constant neutron densities, which are 
equal to the real ones far from the nuclear matter -- neutron gas 
interface, 
reproduces actual numbers of neutrons. The thickness of the 
neutron skin is then defined as  $R_n-R_p$, $R_p$ being defined 
in a similar manner as $R_n$ . 
The
nuclear surface energy term, $E_{{\rm N,surf}}$, gives the
contribution of the interface between  neutron gas and  nuclear
matter; it includes contribution of neutron skin
\cite{LPRL85,PethRav95,LorenzThesis}. In the case of spherical and
cylindrical interface, $E_{{\rm N,surf}}$ includes curvature
correction; the curvature correction vanishes for slabs.

In order to calculate $E_{\rm Coul}$, we used the Wigner-Seitz
approximation. In the case of spheres, bubbles, rods, and tubes,
Wigner-Seitz cells were approximated by spheres, and cylinders, of
radius $R_{\rm cell}$. In the case of slabs, Wigner-Seitz cells
were bounded by planes, with $R_{\rm cell}$ being defined as the
half-distance between plane boundaries of the cell. At given
average  nucleon density, $\rho$, and for an assumed shape of
nuclear structures, the energy density was minimized with respect
to  thermodynamic variables, under the condition of an average
charge neutrality \cite{LPRL85,LorenzThesis,DouchinThesis}. Our
results for $E(\rho)$ for the SLy4 force, together with the energy
density of the uniform $npe$ matter,  are displayed in Fig. 3. For
the sake of convenience, energy densities has been shown after
subtracting the value of the energy density of the bulk nuclear
matter - neutron gas system (charge neutral and in beta
equilibrium), $\Delta E(\rho)\equiv E(\rho)-E_{\rm bulk}(\rho)$.
Spherical nuclei turn out to be energetically preferred over other
nuclear shapes, and also over homogeneous $npe$ matter,  down to
$\rho_{\rm crit}=0.077~{\rm fm^{-3}}$. Therefore, within our set
of possible nuclear shapes, the ground state of neutron-star crust
contains spherical nuclei, down to its bottom edge. 

As was already noticed  in \cite{RavPethW83,Hashimoto84,
LRP93}, exotic 
nuclei will be present only if the filling fraction, $u$, is 
large enough. For a given neutron excess, introducing finite size 
(Coulomb and surface) effects increases the value of  $u$. 
This increase of $u$ is larger when the surface tension is larger, the 
finite-size effect being roughly propotional to it. Therefore, the larger 
the surface tension, the more likely  the exotic nuclei are to appear. 
For all Skyrme forces we have studied so far, our results are in 
agreement with this qualitative rule. In particular, at the proton fraction
$\sim 5-8$\% the surface tension for the SkM$^*$ and SLy forces,  
for which exotic nuclei do not appear,  is much smaller than for 
the Sk1$^\prime$ model, for which exotic nuclei do appear.
 
Actually,
under conditions of thermodynamic equilibrium, transition from the
crust to the uniform liquid takes place at a constant pressure,
and is accompanied by a density jump (first order phase
transition). Using Maxwell construction, we find that the edge of
the crust has density $\rho_{\rm edge}=0.076~{\rm fm^{-3}}$, and
coexists there with uniform $npe$ matter of the density higher by
$\Delta\rho/\rho_{\rm edge}=1.4\%$. Therefore, crust-liquid core
transition is a very weak first-order phase transition; it takes
place at $P_{\rm edge}=0.34~{\rm MeV/fm^3}$.

 Geometrical parameters
characterizing the  bottom layers of the crust are shown in Fig.
4. Here, $R_p$ is the proton radius corresponding to the
equivalent squared proton distribution within nuclei, $R_{\rm
cell}$ is the radius of the Wigner-Seitz cell, and $u$ is the
fraction of volume occupied by nuclear matter (equal to that
occupied by protons). The crust-liquid core transition takes place
not because nuclei grow in size, but rather because they become
closer and closer. The filling fraction $u$ grows rapidly for
$\rho$ approaching $\rho_{\rm edge}$,  from only 5 \% at
$\rho=0.04~{\rm fm^{-3}}$ to nearly 30 \% at $\rho_{\rm edge}$.
Still, less than 30\% of the volume is filled by nuclear matter at
the crust-liquid
 core transition point. 
Our maximum filling factor is significantly larger than the limiting 
value of 0.2  for spherical nuclei phase, derived by Oyamatsu et 
al. \cite{Oyam84} using general considerations involving surface 
and Coulomb energy. This is due to the curvature corrections, 
included in our formula for $E_{\rm N,surf}$ and absent in the 
model of Oyamatsu et al. \cite{Oyam84}. Curvature tension is particularly 
important near the crust-liquid transition point, where the surface tension 
is very small. If we neglect the curvature term in surface energy, we 
find maximum $u$ for nuclei of about 0.2, consistent 
with Oyamatsu et al. \cite{Oyam84}.  
Finally, let us mention that no  proton drip occurs in the ground state of the
crust.

More detailed information on neutron-rich nuclei, present in the
ground state of the bottom layers of neutron-star crust with
$\rho\ga {1\over 3}\rho_0$, can be found in Fig. 5. Number of
nucleons in a nucleus, $A$, grows monotonically, and reaches about
600 at the edge of the crust. However, the number of protons
changes rather weakly, from $Z\simeq 40$ near  neutron drip, to
$Z\simeq 50$ near the edge of the crust. Our results for $Z$ of
spherical nuclei are  similar to those  obtained in \cite{Oyam93,
RBP72}, but are somewhat higher than those obtained using a
relativistic mean-field model in \cite{Sumiy95}. An interesting
quantity is the number of neutrons  forming neutron skin, $N_{\rm
n,surf}$. For $\rho>{1\over 3}\rho_0$, we find that $N_{\rm
n,surf}$ decreases with increasing $\rho$, despite an increase of
nuclear radius. This behavior is due to the fact, that neutron
densities outside and inside nuclei become more an more alike,
while thickness of neutron skin decreases, with increasing $\rho$,
for $\rho>{1\over 3}\rho_0$.

Near  the bottom of the crust, spherical nuclei are very heavy,
$A\sim 300-500$, and doubts concerning their stability with
respect to deformation and fission arise. An approximate condition
for fission in dense neutron-star matter is $R_p/R_{\rm cell}\ga
1/2$ \cite{PethRav95}; this condition is fulfilled near the edge
of the crust, Fig. 4. However, curvature terms in $E_{\rm N,surf}$
and $E_{\rm Coul}$, which were not included  in generalized
Bohr-Wheeler condition of \cite{PethRav95}, as well as shell
correction to $E$ \cite{OyamYam94}, stabilize large spherical
nuclei. Clearly, the problem of stability of spherical nuclei near
the bottom of neutron-star crust deserves further investigation.
In any case, however, forming rod-like nuclei  in the ground state
of neutron-star crust  seems to be excluded  by the energetical
arguments, Fig. 3.

In the CLDM approximation with SLy4 force, uniform $npe$ matter is
present in the ground state of the neutron-star interior at the
density exceeding  $\rho_{\rm liq}=0.077~{\rm fm^{-3}}$. Comparing
the value of $\rho_{\rm liq}$ with $\rho_Q$, we have an
independent test of precision of our CLDM calculations. The
relative  difference between these two quantities is  less than
3\%, which is of the order-of-magnitude of the relative density
jump in the crust - liquid core transition, obtained within the
CLDM. For the SLy4 force, the crust - liquid core transition takes
thus place at half of the normal nuclear density, and very similar
value is obtained for the SLy7 force.

The structure of the ground state of neutron-star interior at
subnuclear densities, determined in this Letter  for the SLy4
force, implies that the bottom layer of the  neutron star crust is
a three-dimensional Coulomb crystal, with spherical nuclei in the
lattice sites, the edge of the crystal being located at $\rho_{\rm
edge}\simeq 0.08~{\rm fm^{-3}}$.

We  combined the SLy4 EOS  of the inner crust with that of the
liquid core, getting in this way a physically consistent, unified
description of the neutron star interior beyond neutron drip. As
for the outer crust, we have chosen  the EOS of Haensel and Pichon
\cite{HP94}, in which maximum use of recent experimental
information on neutron rich nuclei was made. For neutron star with
canonical mass of $1.4~{\rm M}_\odot$, the crust contains $\Delta
M_{\rm crust}/M=1.2\%$  of the total stellar mass, and its
relative contribution to the total moment of inertia is $\Delta
I_{\rm crust}/I=2.6\%$. The corresponding values for the FPS
neutron star model are some 30\% lower, despite larger value of
$\rho_{\rm edge}({\rm FPS})= 0.10~{\rm fm^{-3}}$. It turns out,
however,  that  $P_{\rm edge}({\rm FPS})\simeq P_{\rm edge}({\rm
SLy4})$, because the FPS EOS near $\rho_{\rm edge}$ is
significantly softer than the SLy4 one. The difference in the
$\Delta M_{\rm crust}/M$ for both models can thus be explained by
the difference in the $R^4(1-2GM/Rc^2)$ factor appearing in an
approximate expression for $\Delta M_{\rm crust}$ \cite{LRP93};
the FPS neutron star of $1.4~{\rm M_\odot}$ is more compact than
the SLy4 one. Similar arguments can be used to explain the
difference between the $\Delta I_{\rm crust}/I$ values for the
SLy4 and FPS forces (see \cite{LRP93} for the approximate formula
to be used). However, the most important difference between the
SLy4 and the FPS crusts is that in our case the whole crust, down
to its bottom edge, is a Coulomb crystal, while in the FPS case
the bottom layer, constituting  about half of the crust mass,
contains nonstandard nuclear shapes (rods, plates, tubes,
bubbles), and therefore has elastic and transport properties very
different from those of a three-dimensional Coulomb  crystal
\cite{PethPot98}. Finally, presence of a Coulomb crystal is also
relevant for the crust - liquid core boundary condition in
rotating neutron stars.

\begin{center}
                 {\bf ACKNOWLEDGMENTS}
\end{center}
We are grateful to G. Chabrier and J. Meyer for a critical reading
of the preliminary version of the manuscript and for helpful
remarks. P.H. acknowledges the hospitality of ENS de Lyon  during
his stay with the theoretical astrophysics group of CRAL.
 This work was supported in part by the KBN grant No. 2 P03D
014 13 and by the French CNRS/MAE program Jumelage Astronomie France-Pologne.
During his visits at DARC, P.Haensel was supported by the PAST 
professorship of French MENRT. 


 \pagebreak
\begin{figure}
\begin{center}
\leavevmode
\epsfxsize=12cm
\epsfbox{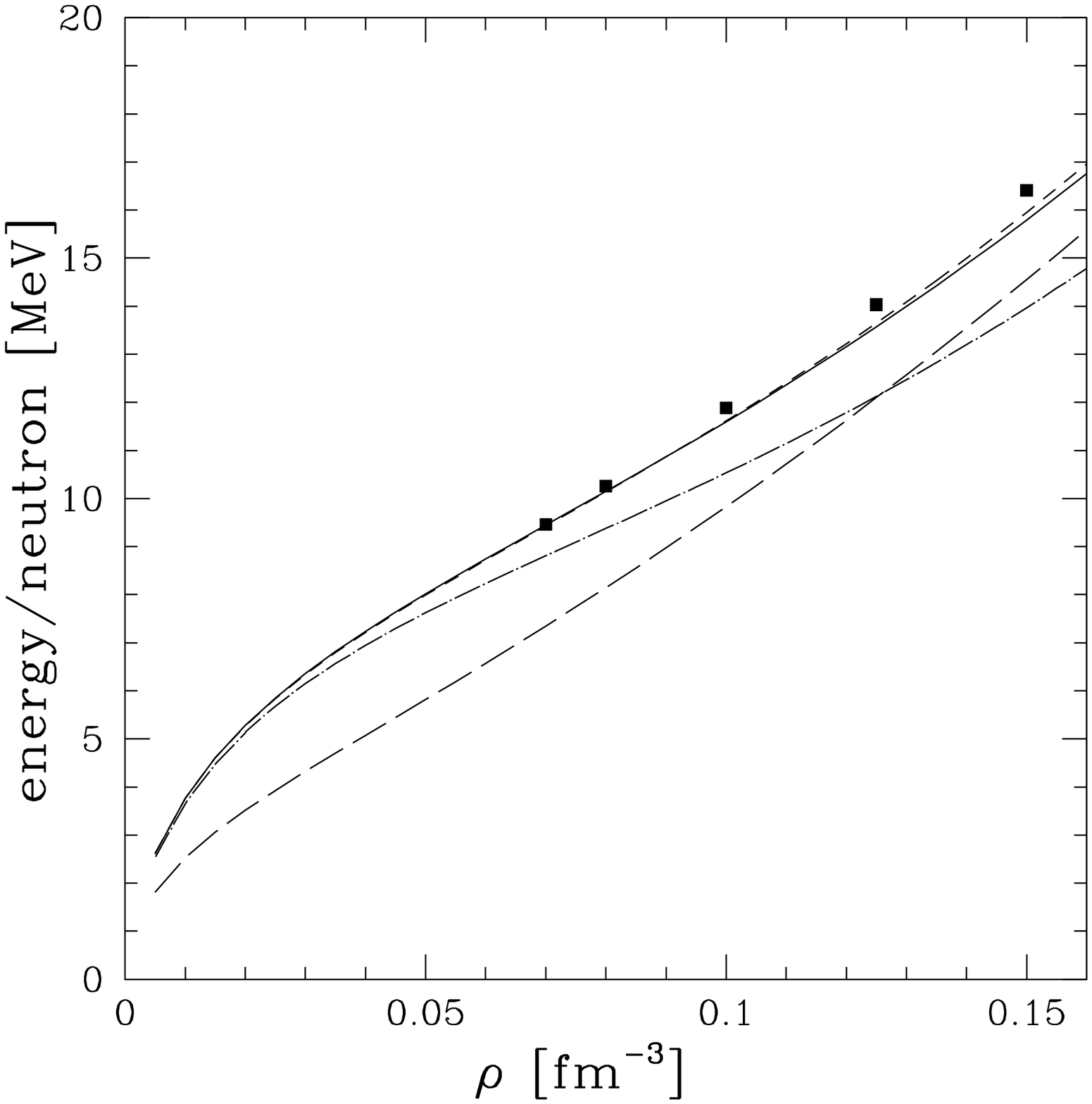}
\end{center}
\caption{} { Energy per neutron versus neutron  density for
neutron matter. Filled squares: results of Wiringa et al.
\cite{WFF88} for the UV14+UVIII model. Solid line: the SLy4 model.
Short dashes: the SLy7 model. For the sake of comparison, we show
also results obtained for the SkM$^*$ (long dashes) and the
Sk1$^\prime$ (long dash-dot line) models. }
\end{figure}
\vfill
\eject
\begin{figure}
\begin{center}
\leavevmode \epsfxsize=16cm \epsfbox{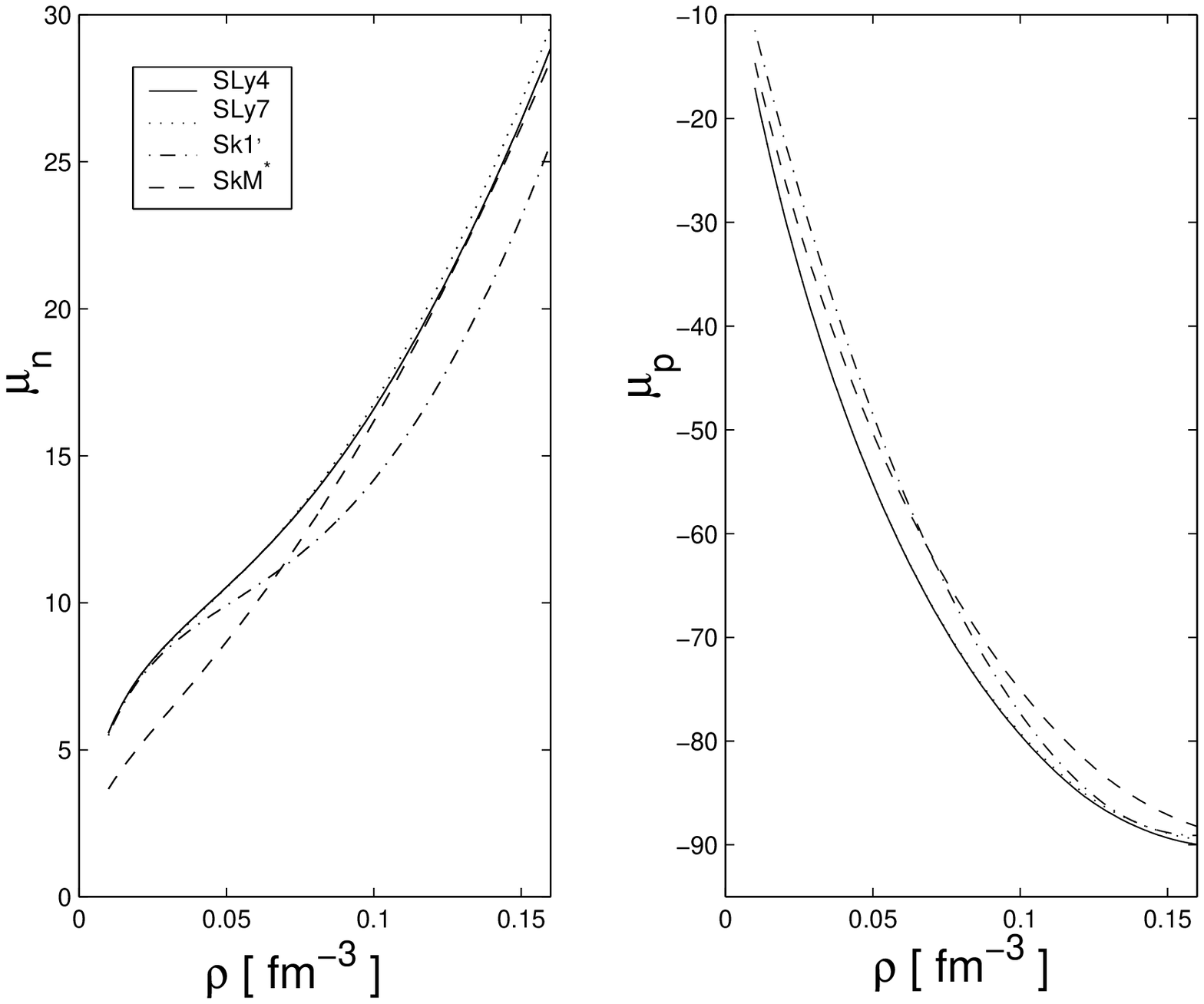}
\end{center}
\caption{} {Neutron and proton chemical potentials (with rest energy 
subtracted), $\mu_n$, $\mu_p$,  versus total nucleon density, 
$\rho$, for the SLy and  SkM$^*$, Sk1$^\prime$ forces, respectively. 
The $\mu_p$ curve for the SLy7 force cannot be graphically distinguished 
from the SLy4 one. }
\end{figure}
\vfill \eject
\begin{figure}
\begin{center}
\leavevmode \epsfxsize=12cm \epsfbox{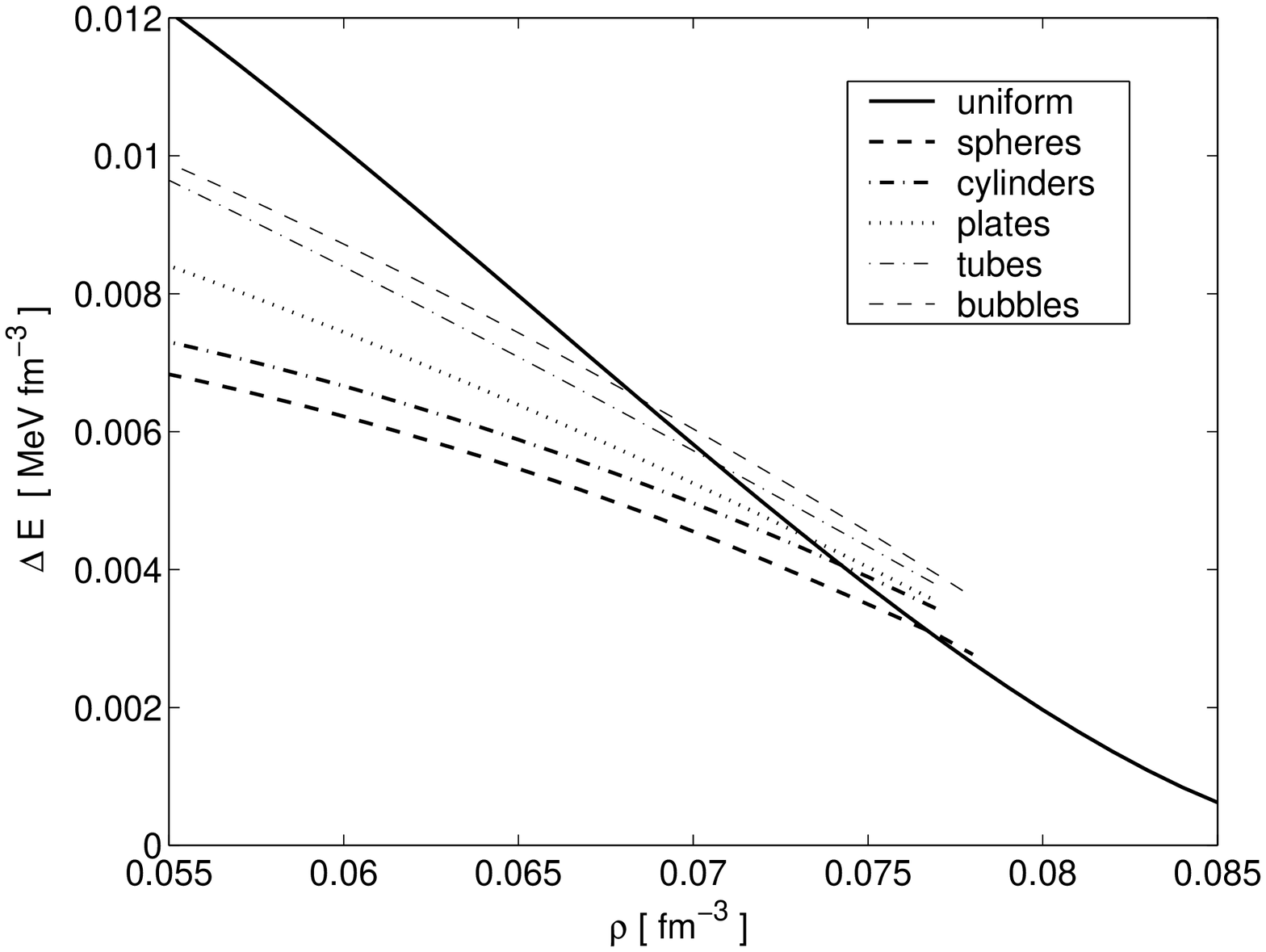}
\end{center}
\caption{} { Energy density of a given phase of inner-crust matter
minus that  of  the bulk nuclear matter - neutron gas system, as a
function of the average nucleon density, $\rho$. Thick solid line
corresponds to the homogeneous $npe$ matter. Calculations
performed for the SLy4 force. }
\end{figure}
\vfill \eject
\begin{figure}
\begin{center}
\leavevmode \epsfxsize=12cm \epsfbox{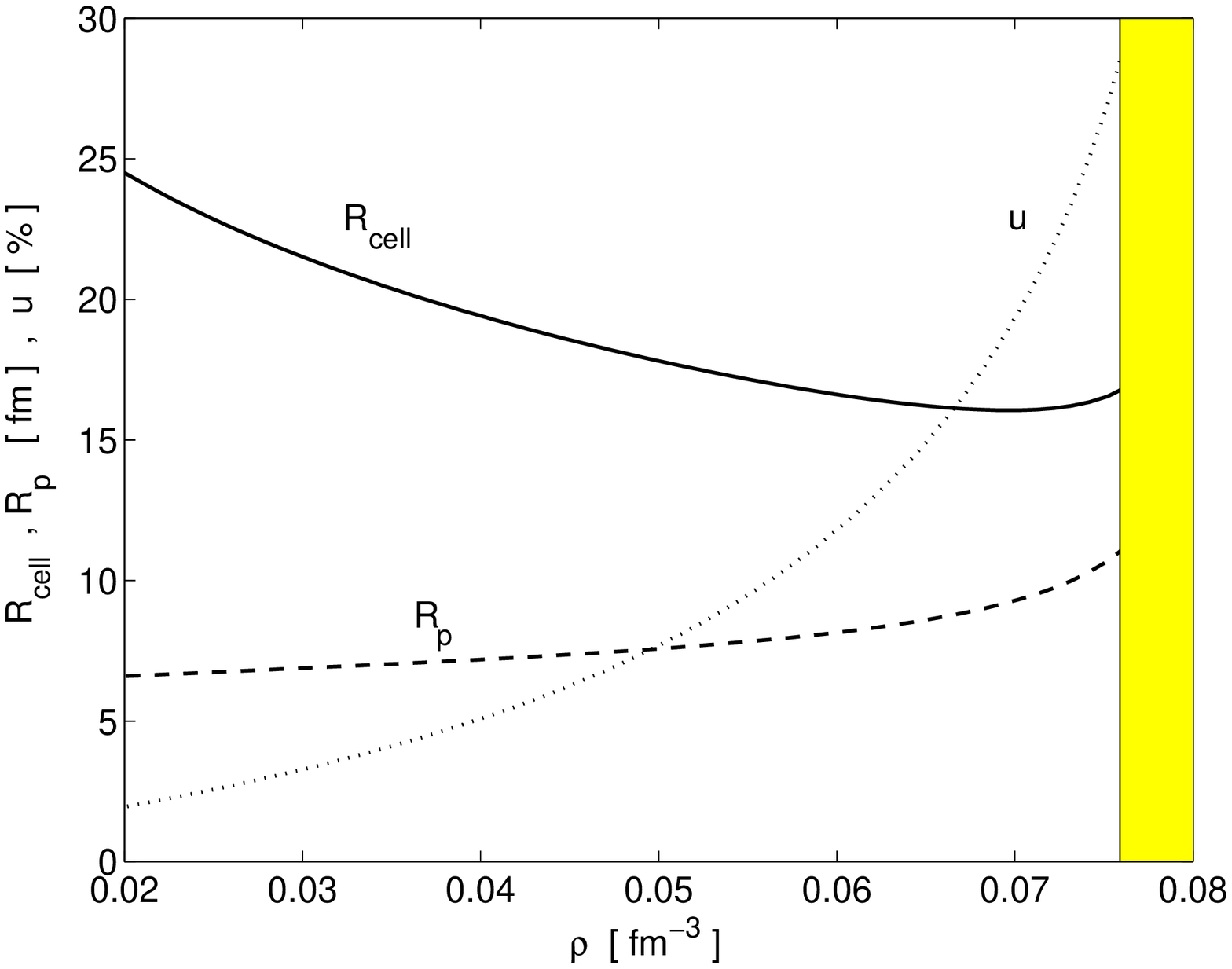}
\end{center}
\caption{} { Radius of  spherical Wigner-Seitz cell, $R_{\rm
cell}$, the proton radius of  spherical nuclei, $R_p$, and
fraction of volume filled by nuclear matter, $u$ (in percent),
versus average nucleon density, $\rho$. Shaded area corresponds to
the homogeneous  $npe$ matter. Calculations performed for the SLy4
force.}
\end{figure}
\vfill \eject
\begin{figure}
\begin{center}
\leavevmode \epsfxsize=12cm \epsfbox{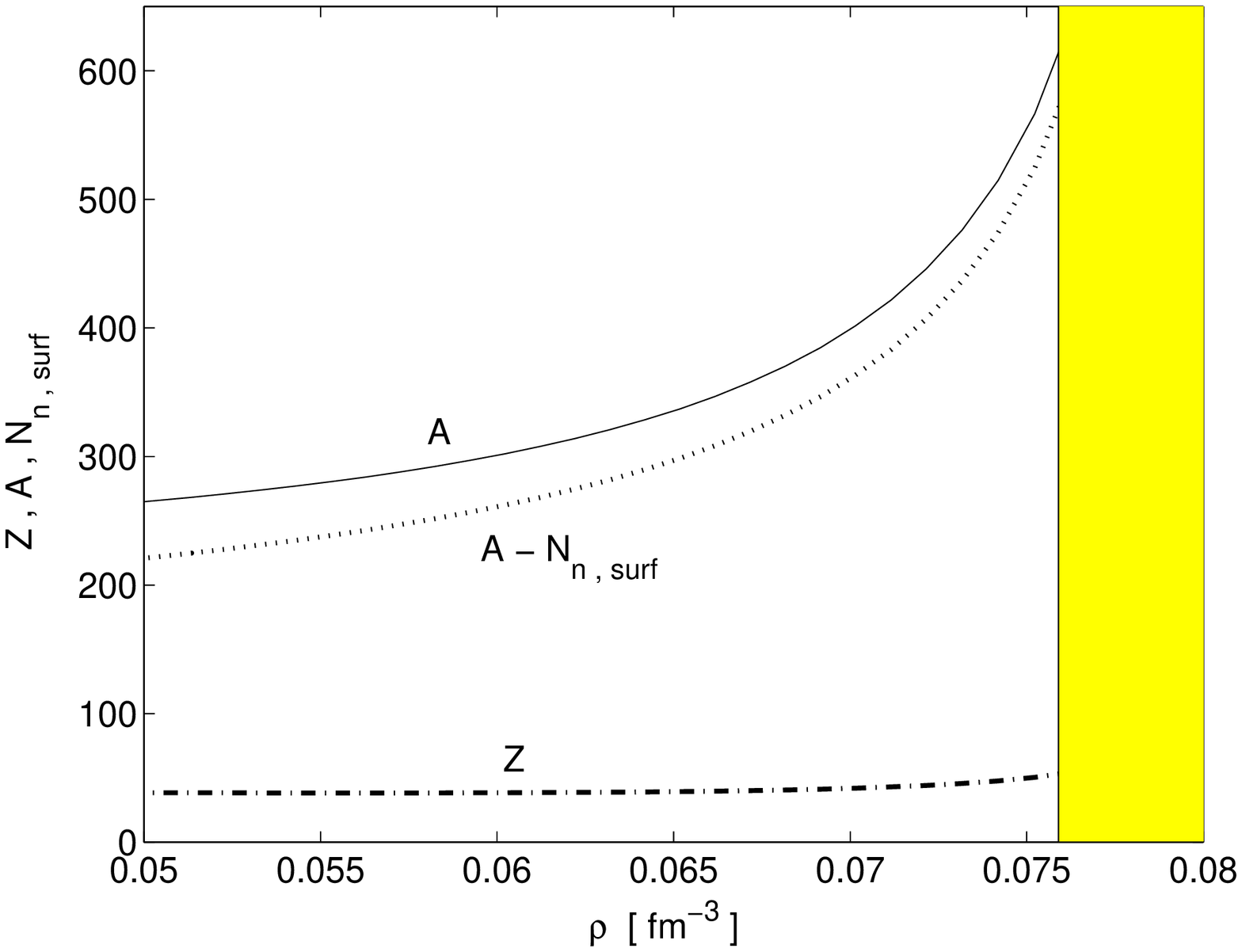}
\end{center}
\caption{} { Mass number of spherical nuclei, $A$, and their
proton number, $Z$, versus average nucleon density, $\rho$. Dotted
line corresponds to number of nucleons after deducing neutrons
belonging to  neutron skin. Shaded area corresponds to the
homogeneous  $npe$ matter. Calculations performed for the SLy4
force.}
\end{figure}
\vfill \eject
\begin{tabular}{ccccc}
\multicolumn{5}{c}{Table 1}\\
&&&&\\
\multicolumn{5}{c}{Parameter values of the Skyrme forces}\\
&&&&\\
\hline\hline
&&&&\\
force & SLy4 &  SLy7  &  SkM$^*$ & Sk1$^\prime$\\
&&&&\\
\hline
&&&&\\
 $t_0$~ $({\rm MeV~fm^3})$ & -2488.91   & -2482.41    & -2645.0 & -1057.3    \\
&&&&\\
$t_1$~ $({\rm MeV~fm^5})$ & 486.82    &  457.97   & 410.0   & 235.9     \\
&&&&\\
$t_2$~$({\rm MeV~fm^5})$ & -546.39      &  -419.85    & -135.0 &   -100.00   \\
&&&&\\
$t_3$~$({\rm MeV~{fm}^{3+3\sigma}})$ & 13777.0& 13677.0  & 15595.0  & 14463.5 \\
&&&&\\
 $\sigma$  &  $1\over 6$ & $1\over 6$  &  $1\over 6$  &  1 \\
&&&&\\
 $x_0$  & 0.834   & 0.846  & 0.09  &  0.2885 \\
&&&&\\
 $x_1$  & -0.344  & -0.511   & 0  & 0  \\
&&&&\\
 $x_2$  &  -1.000 & -1.000  &  0  & 0  \\
&&&&\\
 $x_3$  & 1.354   & 1.391  &  0  &  0.2257 \\
&&&&\\
$W_0$~$({\rm MeV~fm^5})$ &  123.0  & 126.0   & 130.0   & 120.0   \\
&&&&\\
\hline
\end{tabular}
\vfill
\end{document}